\documentclass[twocolumn,aps,pra,showpacs,preprintnumbers,amsmath,amssymb]{revtex4-1}
\usepackage{natbib}
\usepackage{graphicx}
\usepackage{bm}
\usepackage{gensymb}
\usepackage{color}
\begin{document}
\title{Study of maximal bipartite entanglement and robustness in resonating-valence-bond states}
\author{Muzaffar Q. Lone$^1$} 
\author {Sudhakar Yarlagadda$^{1,2}$}
\affiliation{$^1TCMP$ Div. and $^2CAMCS$\\
1/AF Salt Lake, Saha Institute of Nuclear physics, Calcutta, India -700064}

\begin{abstract} 
We study maximal bipartite entanglement in valence-bond states and show that the
 average bipartite entanglement $E_v^2$, between a sub-system of two spins and the rest of the system,  
can be maximized through
a homogenized superposition
of the
valence-bond states.
Our derived maximal $E_v^2$
rapidly increases with system size and 
 saturates at its maximum allowed value.
We also demonstrate that our maximal
 $E^2_v$ states 
are  ground states of 
 an infinite range Heisenberg model (IRHM)
and represent a new class of resonating-valence-bond (RVB) states.
The entangled RVB states produced from our IRHM are robust against interaction of spins with both local 
and global phonons and represent a new class of decoherence free states.
\end{abstract}
\maketitle
\section{Introduction.}
Quantum entanglement, a manifestation of non-locality, 
 is a precious resource for quantum computation and quantum information \cite{6} and 
signifies correlations in many-body systems.
Quantum algorithms that would significantly accelerate  a classical computation
must rely on highly entangled states since  slightly entangled states
can be simulated efficiently on a classical computer \cite{vidal}.
The strength of correlations of fluctuations of observables
(such as density, magnetization, etc.) in a many-body system
is a reflection of the degree of entanglement (for pure states) \cite{latorre}.
Thus, characterization of multi-particle entanglement and production
of maximal/high
 multi-qubit entanglement
is vital for 
quantum computational studies and for mutual enrichment of
 quantum information and many-body condensed matter physics.

Intensive work on entanglement during the 
past decade has led to the proposal of numerous measures of entanglement 
\cite{horodecki,7}.
 While two-party entanglement
is quite well understood, entanglement in a multi-party system
 is an area of immense current 
interest \cite{Meyer,Scott,Mintert,Florio,Adesso,FloriaRapid,20,plastino}.
In the quest for maximally entangled states, so far only few-qubit maximally entangled states
such as two-qubit Bell states, three-qubit 
Greenberger-Horne-Zeilinger (GHZ) states,
 and four-qubit Higuchi-Sudbery (HS) states \cite{HS}
 have been clearly identified.
 It would be of considerable interest
to generate these maximally entangled states as the ground states of
a physically realizable spin model.
While the GHZ
 states could  be obtained as the ground state of
an anisotropic Heisenberg model
 \cite{loss1},
 maximally entangled
four-qubit and five-qubit states could be obtained only as excited states \cite{florio}.

Decoherence is one of the main obstacles for the preparation, observation, and
implementation of multi-qubit entangled states. Since coupling
to the environment and the concomitant entanglement fragility are 
ubiquitous
 \cite{6,schloss}, it is imperative that progress be made in
understanding decoherence as well. In the past decoherence-free-subspace (DFS)
\cite{palma,zanardi} has been shown to exist in the Hilbert space of
a model 
where all qubits of the quantum system 
are coupled to a common environment
with equal strength. This situation manifests
when the distance between
the qubits is negligible compared to the correlation length
of the environment.
If the dynamic symmetry of the system-environment interaction selects a set of
orthonormal basis vectors (of the reduced Hilbert space)
that is unaffected by environmental interaction, then such a subspace is
called a DFS.
An important application of 
decoherence free subspaces
 lies in  
developing  quantum error correcting codes \cite{whaley1,whaley2}.
These subspaces
prevent the loss of information due to destructive environmental interactions
and thus circumvent the need for stabilization methods for
quantum computation and quantum information. Decoherence free
states have also been shown to be
useful for quantum communication between parties 
without a common reference frame
\cite{boileau,chen,bartlett,cabello1,cabello2}.
\\

RVB states
have provided interesting insights for understanding strongly correlated phenomena 
such as physics of high $T_c$ cuprates \cite{anderson,lee},
superconductivity in organic solids \cite{ishiguro}, insulator-superconductor 
transition in boron-doped diamond \cite{ekimov}, etc. Furthermore, RVB states have also
been proposed as robust basis states for topological quantum computation \cite{kitaev}.
The 
multi-particle entanglement has also been investigated in  
RVB states
that were proposed as states close in energy to the ground state
 of the Heisenberg Hamiltonian \cite{LA,sen,alet}. However,  there has been no explicit construction of
 RVB states that would  represent
maximally entangled valence bond states. 
In this paper
 we construct
a new class of RVB states that are ground states of IRHM,
 that have high $E^2_v$ entanglement, and are  decoherence free.
Our IRHM Hamiltonian couples to an
 environment that distinguishes between the qubits
unlike the case considered in Ref. \onlinecite{zanardi}. 
We believe that an
improved understanding of
the entanglement and decoherence properties of RVB states 
will enable their implementation
for quantum computation and quantum information purposes.

 The remainder of this paper is organized as follows. 
In section II, we introduce entanglement entropy and demonstrate explicitly that isotropy and
 homogeneity  
 maximizes entanglement $E^2_v$ between two spins and the rest of the system in the two limits 
of even-N-qubits, i.e., for
$N=4$ and
 $N \rightarrow \infty$.
 For intermediate even-N-spin states,
  our proposed entangled states are only shown to  maximize $E^2_v$
 among valence-bond states. 
 In section III, from the ground states
of our IRHM, we construct explicitly states with maximal $E^2_v$
entanglement among valence-bond 
 states and show that these states  form a special class of RVB states.
Next, in section IV, we analyze the robustness of this special class of entangled states.
   We show that our IRHM Hamiltonian, even upon inclusion of both local and global optical
 phonons that are coupled to spins,
produces ground states that are decoherence free.
Lastly, in the final section V, we conclude after commenting on the physical
realizability of our proposed high $E^2_v$ entangled states.

\section{Entanglement for two-qubit reduced density matrix in isotropic system.}
\begin{figure}
\begin{center}
\includegraphics[width=3.0in,height=2.0in]{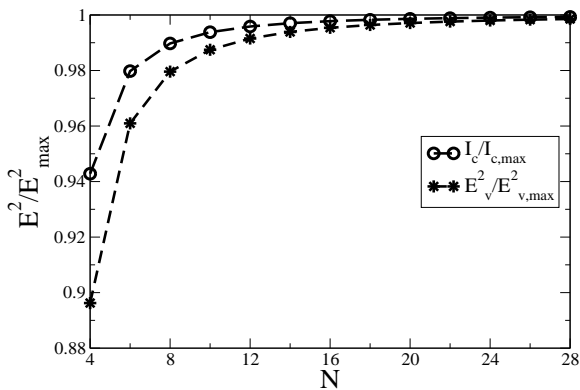}
\caption{Normalized entanglement ${\rm E^2/E^2_{ max}}$, for two-qubit reduced density matrix, measured
by 
 (a) von Neumann entropy (${\rm E^2_{ v}/E^2_{v, max}}$) and (b) i-concurrence (${\rm I_c/I_{c, max}}$), 
for
 N-qubit
 valence-bond systems. }
\label{e_v}
\end{center}
\end{figure}

For a bipartite system AB in a pure state,
von Neumann entropy $E_v$ measures the entanglement between the subsystems A
and B.
From the reduced density matrices 
 $\rho_A \equiv tr_B {\rho^{AB} }$ and $ \rho_B \equiv tr_A {\rho^{AB} }$
obtained from the pure state $\rho^{AB}$, we obtain
\begin{eqnarray}
\!\!\!\! E_v(\rho_{A}) &=& -tr(\rho_A \log_2 \rho_A)
\nonumber \\ 
                       &=& -tr(\rho_B \log_2 \rho_B) = E_v(\rho_B).
\end{eqnarray}
Using the basis $|\downarrow\rangle$, $|\uparrow\rangle$, and 
$ S^i=\frac{1}{2}\sigma^i $, the single qubit density matrix can be written as \cite{21}
\begin{eqnarray}
\rho_i= \left[
\begin{array}{cc}
\frac{1}{2}-\langle S^z_i \rangle & \langle S^{+}_i \rangle\\

\langle S^{-}_i \rangle  & \frac{1}{2}+\langle S^z_i \rangle\\
\end{array}
\right] .
\end{eqnarray}
Throughout this paper, we consider only states $|\Psi \rangle$ that are eigenstates of the
z-component of the total spin operator ($S^z_{Total}$) with eigenvalue $S^z_T$; furthermore,
 we focus on only isotropic states.
It then follows that, $\langle S^z_i \rangle =0$ and $ \langle S^{+}_i \rangle=0$ leading
to the single-qubit density
 matrix to be maximally mixed and thus  maximizing entanglement $E_v(\rho_i)$. 
On realizing that $\langle S^z_i \rangle =0$, we obtain the following expression for the 
two-qubit reduced density matrix \cite{21}:
\begin{eqnarray}
\!\! \rho_{ij} \!= \!\left[\!
\begin{array}{cccc}
\frac{1}{4} + \langle S^z_i S^z_j \rangle  & 0 & 0 & 0\\
0 & \!\frac{1}{4} - \langle S^z_i S^z_j \rangle \! & \langle S^+_i S^-_j\rangle   & 0\\
0 &  \langle S^-_i S^+_j\rangle  & \!\frac{1}{4}- \langle S^z_i S^z_j \rangle \!& 0\\
0 & 0 & 0 & \!\frac{1}{4} + \langle S^z_i S^z_j \rangle \!\\ .
\end{array}
\!\right] . 
\nonumber \\
\label{d_mat}
\end{eqnarray}
Here, isotropy implies 
$0.5 \langle S^-_i S^+_j \rangle = 0.5 \langle S^+_i S^-_j \rangle =\langle S^x_i S^x_j \rangle = \langle S^y_i S^y_j \rangle = \langle S^z_i S^z_j \rangle $.
Thus, the von Neumann entropy 
$ E_v(\rho_{ij}) $
 can be expressed as
\begin{eqnarray}\label{Eij}
\!\!\!\!\!\!\! E_v(\rho_{ij}) = 2 - \frac{1}{4} \!\!&[& \!\!3 (1+
4 \langle S^z_i S^z_j \rangle 
)\log_2(1+
4 \langle S^z_i S^z_j \rangle 
) 
\nonumber \\
&+& \!(1-
12 \langle S^z_i S^z_j \rangle 
 ) \log_2(1-
12 \langle S^z_i S^z_j \rangle 
  )
] .                              
\end{eqnarray} 
For our states, $S^z_{Total}|\Psi \rangle = 0$ which implies
that $\langle S^{z}_i \sum_j S_j^{z}\rangle=0$, that is,
\begin{eqnarray}
\sum_{j \neq i}\langle S^{z}_{i} S^{z}_{j}\rangle=-\langle {S^{z}_i}^2 \rangle=-\frac{1}{4} .
\label{Sz_const}
\end{eqnarray}
We will now maximize the total entanglement entropy
$\sum_{i,j \neq i} E_v(\rho_{ij})$ subject to the above constraint in Eq. (\ref{Sz_const}).
To this end, we will employ the method of Lagrange multipliers and define the Lagrange
function $\Lambda$ as follows:
\begin{eqnarray}
\Lambda = \sum_{i,j \neq i} E_v(\rho_{ij}) - \sum_{i} 
\lambda_i \left ( \sum_{j \neq i}\langle S^{z}_{i} S^{z}_{j}\rangle+\frac{1}{4} \right ) .
\label{Lambda}
\end{eqnarray}
Then, setting $\frac{\partial \Lambda}{\partial \langle S^{z}_l S_m^{z}\rangle} =0$
yields 
\begin{eqnarray}
\lambda_l = 3 \log_2 \left [\frac{(1-12 \langle S^{z}_l S_m^{z}\rangle)}{(1+4  \langle S^{z}_l S_m^{z}\rangle)}
\right ] ,
\label{lambda_i}
\end{eqnarray}
which implies that the optimal $\langle S^{z}_{i} S^{z}_{j}\rangle $ is independent of $j$
for $j \neq i$.
Consequently, 
 it follows from Eq. (\ref{Sz_const}) that $\sum_{i,j \neq i} E_v(\rho_{ij})$ is maximized
when $\langle S^{z}_{i} S^{z}_{j}\rangle=-\frac{1}{4(N-1)}$,
i.e., when the isotropic state is   
 {\em homogeneous}.
The
average entanglement entropy 
between the subsystem of two spins 
 and the rest
of the system (of $N-2$ spins) $E^2_v \equiv [1/N(N-1)] \sum_{i,j \neq i} E_v(\rho_{ij}) $
has a maximum value given by
\begin{eqnarray}\label{E_vN}
E_{v,max}^2 &=&  -3\left( \frac{1}{4} - \frac{1}{4(N-1)} \right)
\log_2\left( \frac{1}{4} - \frac{1}{4(N-1)} \right)  \nonumber\\
&-&\left(\frac{1}{4} + \frac{3}{4(N-1)} \right)
\log_2\left(\frac{1}{4} +\frac{3}{4(N-1)} \right) . 
\end{eqnarray}
It is interesting to note that for $N \rightarrow \infty$, 
the above expression yields $E_v^2 \rightarrow 2$.
In fact, $E_v^2$ approaches the maximum possible value $2$ quite rapidly as can be seen
from Fig.~\ref{e_v}.
Next, we observe that for $N=4$, our expression for $E_{v,max}^2$ in Eq. (\ref{E_vN})
yields the same entanglement
entropy value $1+0.5 \log_2 3$ as that obtained for the four-qubit maximally entangled
HS state \cite{plastino,HS}.
Furthermore, our approach explains
why all the pairs in the HS state give the same entanglement value.
Contrastingly, for the isotropic ground state,
in the  case of a Heisenberg chain with nearest-neighbor
 interaction, for four spins
 the entanglement  entropy $E^2_v =1.21$ (with $\langle S^z_i S^z_j \rangle= -0.5/3 $)
 while for an infinite chain  the von Neumann entropy $E^2_v = 1.37$ 
(with $\langle S^z_i S^z_j \rangle \approx -0.443/3 $) 
both of which  are far less than our maximal $E^2_v$ values above. 
It is of interest to note that, when
$N=4$ or $N \rightarrow \infty$, the maximal values of $E^2_v$ for isotropic systems
are the same as the maximal $E^2_v$ values for general systems (i.e., systems that
can be either isotropic or non-isotropic).

We also note that homogeneity, under the constraint of Eq. (\ref{Sz_const}),
 maximizes the i-concurrence ($I_c$) \cite{rungta} given by
\begin{eqnarray}
 I_c = \frac{2}{N(N-1)}\sum_{i,j> i} \sqrt{2[1-Tr(\rho_{ij}^2)]} .
\end{eqnarray}
As shown in Fig. 1, $I_c$ also monotonically increases with system size.
Although we considered von Neumann entropy and i-concurrence
as entanglement measures, our 
 homogeneous states
 should also maximize other measures of entanglement for $\rho_{ij}$ in valence-bond systems.

\section{Multi-qubit entangled states.}
\subsection{Highly entangled ground states of IRHM}
We will now demonstrate that 
ground states of
the IRHM Hamiltonian  
\begin{eqnarray}
\!\!\!\!\!\!\!\! H_{\rm IRHM} = J\! \sum_{i,j>i} \!\! \vec{S_i}.\vec{S_j} 
= \frac{J}{2} \! \left [ \!
 \left ( \sum_{i} \vec{S_i} \right )^2 \!\!
 - \! \left ( \sum_{i} \vec{S_i}^2 \right )
 \!\right ] ,
\label{H_gen}
\end{eqnarray} 
will produce the same amount of entanglement as given by Eq. (\ref{E_vN}). We note that $H_{\rm IRHM}$
 commutes with both  $S^z_{Total}$ and $\left ( \sum_{i} \vec{S_i} \right )^2$ ($\equiv S^2_{Total}$).
In Eq. (\ref{H_gen}), it is understood that $J = J^{\star}/(N-1)$ 
(where $ J^{\star}$ is a finite
quantity) so that the energy per site remains finite as $N \rightarrow \infty$.
The eigenstates of $H_{\rm IRHM}$
 correspond to eigenenergies
\begin{eqnarray}
 E_{S_T} = \frac{J}{2} \left [ S_T (S_T + 1) - \frac{3N}{4} \right ], 
\end{eqnarray}
where $S_T$ is the total spin eigenvalue. The ground state corresponds to $S_T=0$ which is rotationally
invariant and also implies that $S^z_{T}=0$.
 Thus for a homogenized linear combination of the $S_T =0$ states of $H_{\rm IRHM}$, 
 the entanglement is given by Eq. (\ref{E_vN}). It is important to note that, while all
possible $S_T =0 $ states are ground states of IRHM, for some Hamiltonians only
some of the possible $S_T=0$ states are ground states.
For instance,
for the nearest-neighbor antiferromagnetic Heisenberg Hamiltonian 
or for the Majumdar-Ghosh model Hamiltonian \cite{ckm} only
some of the possible $S_T=0$ states are ground states and consequently (for these
systems) it is not
 possible to construct a homogenized linear
combination of $S_T=0$ ground states that produces maximal possible 
entanglement $E^2_{v,max}$ given by Eq. (\ref{E_vN}).

We will now proceed to construct  entangled
states for $N$ spins that maximize $E^2_v$.
We first note that there are 
 $(N-1)!!$ eigenstates
with $S_T =0$ for the Hamiltonian of Eq. (\ref{H_gen}).
Each of the $S_T =0$ states
is a product of
$N/2$ two-spin singlet states of the form
{\nolinebreak$|\uparrow\downarrow\rangle-|\downarrow\uparrow\rangle$} (with no pair of singlets sharing a spin).
Of these
$(N-1)!!$ 
 product combinations with $S_T =0$,
 only ${^N}C_{\frac{N}{2}}-{^N}C_{\frac{N}{2} - 1} = N!/[(N/2)!(N/2+1)!]$ 
products
are linearly independent. A particular set of linearly independent $S_T=0$ states are the Rumer states 
\cite{rumer,pauling} which are made up of non-crossing singlets. For $N=6$, the
 Rumer states are the 5 diagrams with non-crossing singlets shown in Fig. 2.
 
For $N=4$ spins,  two linearly independent
$S_T =0$  eigenstates are
$|\Phi^{S_{12} =0}_{12}\rangle \otimes |\Phi^{S_{34} =0}_{34}\rangle$ and
$ |\Phi^{S_{14} =0}_{14}\rangle \otimes |\Phi^{S_{23} =0}_{23}\rangle$
where $|\Phi^{S_{ij} =0}_{ij}\rangle $ is a two-spin singlet state for spins
at sites $i$ and $j$ with $S_{ij}$ being the total spin of $S_i$ and $S_j$. It is worth noting
that
\begin{eqnarray}
|\Phi^{S_{13} =0}_{13}\rangle \otimes |\Phi^{S_{24} =0}_{24}\rangle &=&
 |\Phi^{S_{12} =0}_{12}\rangle \otimes |\Phi^{S_{34} =0}_{34}\rangle
\nonumber \\
&& +
 |\Phi^{S_{14} =0}_{14}\rangle \otimes |\Phi^{S_{23} =0}_{23}\rangle  .
\label{lin_dep}
\end{eqnarray}
Using the above relation one can establish that there are
only ${^N}C_{\frac{N}{2}}-{^N}C_{\frac{N}{2} - 1}$ linearly independent
$S_T =0$ states.
Furthermore, we also observe that 
\begin{eqnarray*}
\!\!\!\![\vec{S_1}.\vec{S_3}+\vec{S_2}.\vec{S_4}+\vec{S_1}.\vec{S_4}+\vec{S_2}.\vec{S_3}] 
|\Phi^{S_{12} =0}_{12}\rangle \otimes |\Phi^{S_{34} =0}_{34}\rangle = 0 ,
\label{}
\end{eqnarray*}
using which it can be shown that any $S_T =0$ state made up of $N/2$ singlets will be an eigenstate
of Eq. (\ref{H_gen}). 

Now,
while the two symmetry broken $S_T=0$ singlet states [on the right hand side of Eq. (\ref{lin_dep})]
are isotropic, they are not homogeneous.
To make our entangled state homogeneous, we construct the
following state that has the symmetry that
 $\langle S_i^{\eta} 
S_j^{\eta} \rangle$
is the same ($=-1/12$) for all pairs $i$ and $j$
 and any $\eta$ ($=x,y,z$):
\begin{eqnarray}
&&
\!\!\!\!\!\!  \omega_3 (|\Phi^{S_{12} =0}_{12}\rangle
 \otimes |\Phi^{S_{34} =0}_{34}\rangle) 
+\omega_3^2(|\Phi^{S_{14} =0}_{14}\rangle
 \otimes |\Phi^{S_{23} =0}_{23}\rangle) 
\nonumber \\ 
  &=& \frac{1}{\sqrt{6}} \left [
 (|\uparrow\downarrow\uparrow\downarrow \rangle +
 |\downarrow\uparrow\downarrow\uparrow \rangle) 
+ \omega_3(|\uparrow\downarrow\downarrow\uparrow\rangle +
|\downarrow\uparrow\uparrow\downarrow\rangle ) 
\right .
\nonumber\\
&&~~~~~~~~
+ \omega_3^2 
\left .
 (|\uparrow\uparrow\downarrow\downarrow \rangle
 +|\downarrow\downarrow\uparrow\uparrow \rangle) \right ] 
\nonumber\\
&&
\!\!\!\!\!\!  
 \equiv |\Psi_{\rm HS}\rangle ,
\label{psi_hs}
\end{eqnarray}
where $\omega_3$'s  are cube roots of unity.
The above 
 $|\Psi_{\rm HS}\rangle $
state is the maximally entangled four-qubit HS state studied by many authors
\cite{HS,plastino,20,florio}.
However, it has not been shown earlier that
 $|\Psi_{\rm HS}\rangle $ can be expressed as a superposition
 of $S_T =0$ states.
Importantly, the half-filled
large $U/t$ Hubbard model on a regular tetrahedron  yields
the above two broken-symmetry $S_T =0$ states as
ground states from which the HS-state
can be constructed. In previous reported works \cite{gu,ibose}, although the authors
studied entanglement in the model $H = J\! \sum_{i,j=1,2,3,4}\vec{S_i}.\vec{S_j} $,
they did not point out that the maximally entangled HS state is a ground state of the model. 

\begin{figure}[]
\begin{center}
\includegraphics[width=2.5in,height=2.0in]{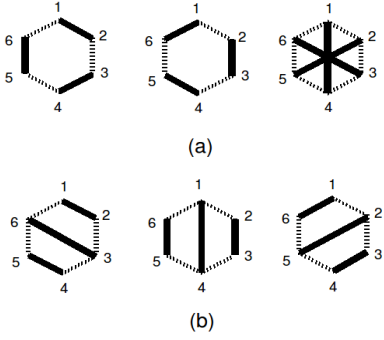}
\caption{Homogenized linear combination of the $S_T=0$ states in both (a) and (b) give the 
same maximal entanglement for six qubits.}
\label{six}
\end{center}
\end{figure}

The above strategy of constructing homogeneous and maximal $E^2_v$ states from isotropic
 four-qubit states can be extended to the six-qubit case as well
using the $S_T=0$ ground states of $H_{\rm IRHM}$ shown in the six
 diagrams of Fig.~\ref{six}.
The entanglement $E^2_v$ for the six-qubit 
 states is maximized by taking suitable
 resonance hybrids  of the diagrams shown in Figs.~\ref{six} (a) or (b).
These resonance states homogenize the ground state in the sense 
mentioned above (i.e., produce site independent correlation functions 
$\langle S_i^{\eta} S_j^{\eta} \rangle =-1/20$):
\begin{eqnarray}
  |\Psi^a_6\rangle &=&\omega_4(| \Phi^{S_{12} =0}_{12} \rangle \otimes
 |\Phi^{S_{34} =0}_{34} \rangle \otimes
 |\Phi^{S_{65} =0}_{65}\rangle) \nonumber \\
&+& \omega_4^2 (|\Phi^{S_{61} =0}_{61}\rangle \otimes
 |\Phi^{S_{23} =0}_{23} \rangle \otimes
 |\Phi^{S_{45} =0}_{45}\rangle) \nonumber \\
&+&\omega_4^3(|\Phi^{S_{14} =0}_{14} \rangle \otimes |\Phi^{S_{25} =0}_{25}\rangle \otimes |\Phi^{S_{36} =0}_{36}\rangle) ,
\end{eqnarray}
and

\begin{eqnarray}
 |\Psi^b_6\rangle &=& \omega_4 (|\Phi^{S_{12} =0}_{12}\rangle \otimes |\Phi^{S_{36} =0}_{36} \rangle \otimes |\Phi^{S_{45} =0}_{45}\rangle) \nonumber \\
&+& \omega_4^2(| \Phi^{S_{23} =0}_{23} \rangle \otimes |\Phi^{S_{14} =0}_{14} \rangle \otimes |\Phi^{S_{56} =0}_{56}\rangle) \nonumber \\
&+&\omega_4^3(|\Phi^{S_{16} =0}_{16} \rangle \otimes |\Phi^{S_{25} =0}_{25}\rangle \otimes |\Phi^{S_{34} =0}_{34}\rangle) ,
\end{eqnarray}
which can be rewritten as

\begin{widetext}
\begin{eqnarray}
 |\Psi^a_6\rangle &=& \frac{1}{\sqrt{20}}[(|\uparrow\downarrow\uparrow\downarrow\uparrow\downarrow\rangle - |\downarrow\uparrow\downarrow\uparrow\downarrow\uparrow \rangle) +\omega_4(|\uparrow\downarrow\uparrow\downarrow\downarrow\uparrow\rangle  
+ |\uparrow\downarrow\downarrow\uparrow\uparrow\downarrow\rangle + |\downarrow\uparrow\uparrow\downarrow\uparrow\downarrow \rangle -|\uparrow\downarrow\downarrow\uparrow\downarrow\uparrow\rangle -|\downarrow\uparrow\uparrow\downarrow\downarrow\uparrow\rangle 
-|\downarrow\uparrow\downarrow\uparrow\uparrow\downarrow\rangle ) \nonumber \\
&&~~~~~~
+ \omega_4^2(|\uparrow\uparrow\downarrow\downarrow\uparrow\downarrow\rangle+| \uparrow\downarrow\uparrow\uparrow\downarrow\downarrow\rangle +|\downarrow\downarrow\uparrow\downarrow\uparrow\uparrow\rangle 
-|\uparrow\uparrow\downarrow\uparrow\downarrow\downarrow\rangle -|\downarrow\uparrow\downarrow\downarrow\uparrow\uparrow\rangle-|\downarrow\downarrow\uparrow\uparrow\downarrow\uparrow\rangle) \nonumber \\
&&~~~~~~
+\omega_4^3(|\uparrow\uparrow\uparrow\downarrow\downarrow\downarrow\rangle 
+ |\uparrow\downarrow\downarrow\downarrow\uparrow\uparrow\rangle+ |\downarrow\downarrow\uparrow\uparrow\uparrow\downarrow\rangle -|\uparrow\uparrow\downarrow\downarrow\downarrow\uparrow\rangle -|\downarrow\uparrow\uparrow\uparrow\downarrow\downarrow\rangle 
- |\downarrow\downarrow\downarrow\uparrow\uparrow\uparrow\rangle)] ,
\label{psi6a}
\end{eqnarray}
and
 \begin{eqnarray}
 |\Psi^b_6\rangle &=& \frac{1}{\sqrt{20}}[(|\uparrow\downarrow\uparrow\downarrow\uparrow\downarrow\rangle - |\downarrow\uparrow\downarrow\uparrow\downarrow\uparrow \rangle) +\omega_4(|\uparrow\downarrow\uparrow\uparrow\downarrow\downarrow\rangle  
+ |\uparrow\downarrow\downarrow\downarrow\uparrow\uparrow\rangle + |\downarrow\uparrow\uparrow\downarrow\uparrow\downarrow \rangle -|\uparrow\downarrow\downarrow\uparrow\downarrow\uparrow\rangle -|\downarrow\uparrow\uparrow\uparrow\downarrow\downarrow\rangle 
-|\downarrow\uparrow\downarrow\downarrow\uparrow\uparrow\rangle ) \nonumber \\
&&~~~~~~
+ \omega_4^2(|\uparrow\uparrow\downarrow\downarrow\uparrow\downarrow\rangle+| \uparrow\downarrow\uparrow\downarrow\downarrow\uparrow\rangle +|\downarrow\downarrow\uparrow\uparrow\uparrow\downarrow\rangle 
-|\uparrow\uparrow\downarrow\downarrow\downarrow\uparrow\rangle -|\downarrow\uparrow\downarrow\uparrow\uparrow\downarrow\rangle-|\downarrow\downarrow\uparrow\uparrow\downarrow\uparrow\rangle) \nonumber \\
&&~~~~~~
+\omega_4^3(|\uparrow\uparrow\uparrow\downarrow\downarrow\downarrow\rangle
+ |\downarrow\downarrow\uparrow\downarrow\uparrow\uparrow\rangle+ |\uparrow\downarrow\downarrow\uparrow\uparrow\downarrow\rangle -|\uparrow\uparrow\downarrow\uparrow\downarrow\downarrow\rangle -|\downarrow\uparrow\uparrow\downarrow\downarrow\uparrow\rangle 
- |\downarrow\downarrow\downarrow\uparrow\uparrow\uparrow\rangle)] ,
\label{psi6b}
\end{eqnarray}
\end{widetext}
where $\omega_4$'s are fourth roots of unity. The von Neumann entropy $E^2_v$,
for two-qubit reduced density matrix obtained from 
$ |\Psi^a_6\rangle$,
$ (|\Psi^a_6\rangle)^*$,
$ |\Psi^b_6\rangle$, or
$ (|\Psi^b_6\rangle)^*$,
is 1.921964 which is the
same as $E^2_{v,max}$ proposed by our general formula in Eq. (\ref{E_vN}).
 The conjecture made by Brown \emph{et al.} \cite{20}
that the multi-qubit maximally entangled states always have their reduced
 single qubit density matrix maximally mixed is satisfied by our states
 $|\Psi_6\rangle$.  It should be noted that the highly
entangled six-qubit state reported in Ref. \cite{plastino}, although yields
a higher entanglement, is not an eigenstate
of the $S^z_{Total}$ operator.

 Cabello \cite{cab} has constructed supersinglets of four and six 
qubits which are decoherence free. It is interesting to note that they
form the ground states of our IRHM Hamiltonian. However, the average two particle 
von Neumann entanglement entropy $E^2_v$
for these supersinglets
 is less than the entanglement for our states constructed above. 
In the case of the four qubit supersinglet state $\left | \mathcal{S}^{(2)}_4 \right >$, 
 the entanglement entropies for all possible bipartitions 
are given by $ E_v(\rho_{12}) =  E_v(\rho_{34}) = 1.5849$, 
$ E_v(\rho_{13}) =E_v(\rho_{24}) = 1.2075$, $ E_v(\rho_{14}) =E_v(\rho_{23})= 1.2075$
which together yield the average
 two particle entropy value
 $1.3333$, i.e.,
 a quantity clearly smaller than the von Neumann entropy $E^2_v = 1 + 0.5\log_2 3 \approx 1.79248$ 
obtained for the four-qubit maximally entangled HS state.
In the six qubit supersinglet $\left | \mathcal{S}^{(2)}_6 \right >$ case, the various bipartitions
produce the average entropy 
$E^2_v = 1.657996$ which is noticeably smaller than the entropy value $E^2_{v,max} = 1.921964$ 
obtained for our six qubit
entangled states. 

On the issue of how many maximal $E^2_{v}$ states are possible
for $N$ qubit  systems, we would like
to say that it cannot exceed the total number of linearly independent $S_T =0$ states (i.e.,
the total number of Rumer states ). As regards existence of homogeneous states
for $N \ge 8$ qubit systems, we provide arguments in Appendix A. 
\subsection{Resonating-valence-bond picture}
The construction of our entangled maximal $E^2_v$ states from the ground states of IRHM
can be visualized 
by using a 
RVB picture. 
Our maximal $E^2_v$ states can be regarded as a new class of RVB states made of homogenized
superposition of isotropic $S_T =0$ valence bond states.
We will now compare the entanglement properties of our RVB
states and the general RVB states $|\Psi\rangle_{\rm rvb}$ of Ref. \onlinecite{sen}
given below:
\begin{eqnarray}
\!\!\!\!\!\!|\Psi\rangle_{\rm rvb} =\sum_{\substack{i_{\alpha}\in A \\ j_{\beta}\in B}} 
f(i_1,...,i_M,j_1,...,j_M)
|(i_1,j_1)...(i_M,j_M)\rangle ,
\nonumber
\label{rvb}
\end{eqnarray}
where $M$ represents the number of sites in each sub-lattice and $f$ is assumed to be isotropic
 over the lattice. 
Also, $ |(i_k,j_k)\rangle \equiv 
\frac{1}{\sqrt2}(|\frac{1}{2}\rangle_{i_k} |-\frac{1}{2}\rangle_{j_k}
 -|-\frac{1}{2}\rangle_{i_k} |\frac{1}{2}\rangle_{j_k})$ 
denotes the singlet dimer connecting a site in sub-lattice $A$ with a site in sub-lattice $B$.
 The valence bond basis (used for the above RVB state $|\Psi\rangle_{\rm rvb}$) form an over complete set  
while 
our RVB states are constructed from a complete set of ${^N}C_{\frac{N}{2}}-{^N}C_{\frac{N}{2} - 1}$ states.

The rotational invariance of the two qubit reduced density matrix of the RVB states 
allows us to write them in the form of a Werner state:
\begin{eqnarray}
\rho_{w}(p) =p|(ij)\rangle \langle (ij)|+\frac{1-p}{4}I_4 ,
\end{eqnarray}
where for $1/3 < p \leq 1$ the Werner state has the spins at $i$ and $j$ entangled with each other.
For the special case of the ``RVB gas'' for the
$|\Psi\rangle_{\rm rvb}$ state, where $f$ is a constant (corresponding to
equal amplitude superposition of all bipartite valence bond coverings),
one gets the exact result $p=1/3+2/(3N)$ which implies that all finite size systems
have a non-zero tangle (or entanglement) between the two sites \cite{sen}.
 Next, for the ``RVB liquid'' case
(involving equal amplitude superposition of all  nearest-neighbor
singlet valence bond coverings of a lattice), Monte Carlo calculations 
extrapolated to the thermodynamic limit
yield $p= 0.3946(3) > 1/3$ \cite{sen}, i.e., a non-zero tangle
between the two sites \cite{note}.
In contrast, our maximal $E^2_v$ RVB states yield zero entanglement between the two spins
for all system sizes
as demonstrated below.
It has been shown that the $SU(2)$ symmetry of the RVB states
ensures that the two-spin correlation function and the parameter $p$ of the Werner state
are related as
\begin{eqnarray}
\langle\Psi|\vec{S_i}.\vec{S_j}|\Psi\rangle = -\frac{3}{4}p .
\end{eqnarray} 
Then, since our RVB states produce 
\begin{eqnarray}
\langle S^x_iS^x_j \rangle = \langle S^y_iS^y_j \rangle = \langle S^z_iS^z_j \rangle = \frac{-1}{4(N-1)} ,
\label{SS}
\end{eqnarray}
 it follows that $p=\frac{1}{N-1}$ 
and thus for all even $N\ge 4$ systems we get zero entanglement between the two sites.
Lastly, based on Ref. \onlinecite{sen}, we find that the monogamy argument
yields the bound $p \le 1/3 + 2/(3\sqrt{N-1})$ while the quantum telecloning
argument produces the bound $p \le 1/3 + 2/[3 (N-1)]$. Compared to our exact value of $p = 1/(N-1)$,
both these bounds
are weaker bounds (i.e., show zero two-site entanglement
with certainty only in the thermodynamic limit). Thus 
(among various RVB states) we see that our  entangled
high $E^2_v$ RVB states, while producing maximum entanglement between a pair and the rest of the system,
yield zero entanglement among the two spins of the pair.

We will now remark on the entanglement of any set of $n$ ($>2$) sites
 with the rest of the system.  The valence bond entanglement entropy picture
 of Alet {\emph{et al.}} in Ref. \onlinecite{alet} uses the number of valence bond two-spin singlets shared
by the two subsystems as a measure of their entanglement with each other. Thus, our `homogenized' RVB states,
will always have shared bonds between the two subsystems and thus show 
high bipartite entanglement.
However, exact quantitative analysis
needs to be carried out.
Extending our approach 
to deriving the n-qubit reduced density matrix in terms of $n$-particle correlation
functions (when $n >2$) and obtaining entanglement expressions and wavefunctions
 corresponding to maximal bipartite entanglement is a non-trivial exercise and is left for future studies.

\section{Robust entanglement.}
The real quantum computer will not be free from noise
and thus the entangled states have a tendency to undergo decoherence.
 However 
if the entangled state,
in the presence of potentially decoherence producing interactions,
still remains entangled we call such a state robust otherwise it is fragile.
We will now present two extreme cases of
system-environment interaction scenarios where the ground states
of our IRHM as well as the highly entangled RVB states (that we construct from
the ground states) are decoherence free.

\subsection{Decoherence due to local optical phonons}
To study decoherence due to phonons, we consider interaction with optical phonons such as would
be encountered when considering a Hubbard model.
 The total Hamiltonian $H_T$ is given by
\begin{eqnarray}
\!\!\!\! H_T = H_{\rm IRHM}+ g \omega_0 \sum_i S^z_i (a^{\dagger}_i + a_i) + \omega_0 \sum_i a^{\dagger}_i a_i ,
\label{Ham}
\end{eqnarray}
where $a$ is the phonon destruction operator, $\omega_0$ is the optical phonon frequency, and $g$ is the
coupling strength.  Then, using the steps given in Appendix B,
we obtain the following effective Hamiltonian $H_{e}$ through second-order perturbation theory
for strong coupling ($g>1$) and non-adiabatic ($J/\omega_0 \leq 1$) conditions:
\begin{eqnarray}
H_{e} \! &=& \! \sum_{i, j > i } 
 [J_{\perp}({S_i^{x}}{S_j^{x}} + {S_i^{y}}{S_j^{y}}) +
 J_{\parallel} {S_i^{z}}{S_j^{z}}] 
-g^2 \omega_0  \sum_i S^z_i ,
\nonumber \\
\label{He}
\end{eqnarray}
where 
\begin{eqnarray}
J_{\perp} \equiv J e^{-g^2} - (N-2) f_1 (g) \frac{J^2 e^{-2g^2}}{2 \omega_0} ,
 \end{eqnarray}
\begin{eqnarray}
 J_{\parallel} \equiv 
J + [2f_1 (g)+f_2(g)]\frac{J^2 e^{-2g^2}}{{2 \omega_0}} , 
 \end{eqnarray}
 with
$f_1(g) \equiv \sum^{\infty}_{n=1} g^{2n}/(n!n)$
 and
$f_2(g) \equiv \sum^{\infty}_{n=1}\sum^{\infty}_{m=1} g^{2(n+m)}/[n!m!(n+m)]$. 
It is interesting to note
that the eigenstates of the effective Hamiltonian $H_{e}$ in Eq. (\ref{He}) are identical to
those of the original Hamiltonian $H_{\rm IRHM}$ in Eq. (\ref{H_gen}) because $[\sum_{i, j > i } 
 (S_i^{z} S_j^{z}), H_{\rm IRHM}] =0$. Furthermore, even upon carrying out higher-order 
(i.e., beyond second-order) perturbation theory
(as discussed in Appendix B) ,
 we still get an effective Hamiltonian $H_{eff}$ of the following form
that has the same eigenstates as
the IRHM. 
\begin{eqnarray}
H_{eff} \! &=& \! \sum_{i, j > i } 
 \left [ [J_{xy}(\sum_k S_k^z) ({S_i^{x}}{S_j^{x}} + {S_i^{y}}{S_j^{y}}) \right ]
\nonumber \\
 &&+
 \sum_i J_{z} (\sum_k S_k^z)S_i^{z} ,
\label{Heff}
\end{eqnarray}
where $J_{xy}$ and $J_z$ are functions of the $S^z_{Total}$  ($= \sum_k S_k^z$ ) operator.
It is the infinite range of the Heisenberg model that enables  
 the eigenstates of the system to 
remain unchanged.
Furthermore, the ground state can also remain
unchanged (with $S_T =0$ and $S^z_{T} =0$) if the effect
of the term  $ -C \sum_i S^z_i$ with $C >0$ [such as the
last term in Eq. (\ref{He})] is canceled by a magnetic field.
Thus the set of linearly independent ground states form
 the lowest energy subspace  that is immune to decoherence.
Furthermore, our highly entangled RVB states constructed from these
ground states [such as those given by Eq. (\ref{psi_hs}) for four qubits
 and by Eqs. (\ref{psi6a})--(\ref{psi6b}) for six qubits] 
are also free of decoherence.
Next, we study the decoherence in a dynamical context also and see how
such ground  states can
 remain robust 
and constitute a decoherence free subspace.

\subsection{Dynamical evolution and DFS}
The robustness of entanglement in a system can also be looked from the dynamical perspective.
We consider the following total Hamiltonian where all qubits of our IRHM interact identically
with the environment.
\begin{eqnarray}
\!\!\!\! H_{Tot} = H_{\rm IRHM}+ && \sum_i S^z_i \sum_{k}g_k \omega_k (a^{\dagger}_k + a_k) 
\nonumber \\
 && +  \sum_k \omega_k a^{\dagger}_k a_k .
\label{Ham2}
\end{eqnarray}
 The dynamics of the system  
can be studied through the following 
non-Markovian master equation for the reduced density operator $\rho_S(t)$
\cite{YU}:
\begin{eqnarray}
\frac{d \rho_S(t)}{dt}
=&&
 -i[H_S,\rho_S(t)] + F(t)[L\rho_S(t),L] 
\nonumber \\
&&+F^*(t)[L,\rho_S(t) L] , 
\end{eqnarray}
where $H_S$ is the Hamiltonian of the system and $L$ is the system operator
 that couples with the bath and satisfies the constraint $[L,H_S]=0$.
For the  total Hamiltonian in Eq. (\ref{Ham2}), $H_S=H_{\rm IRHM}$ and $L = \sum_i S^z_i$.
 Also, $ F(t)=\int_0^{t} \alpha(t-s)ds$ 
where $\alpha(t-s)= \eta(t-s)+i\nu(t-s)$ is the bath correlation function at temperature $T$ with
\begin{eqnarray}
\eta(t-s) =\sum_{k} |g_{k}|^2 \coth(\frac{\omega_{k}}{2k_B T} )\cos[\omega_{k}(t-s)] , \nonumber \\
\nu(t-s) =-\sum_{k} |g_{k}|^2 \sin[\omega_{k}(t-s)] .
\end{eqnarray}
The function  $F(t)$ 
governs the non-Markovian dynamical features of the system.

Let $\{| n \rangle \}$ be the eigen basis  in which both the operators $L$ and $H_S$ are simultaneously
diagonalized. Upon solving the master equation explicitly we get \cite{YU}:
\begin{eqnarray}
\rho_{mn} (t) =&& \langle n|\rho_S|m\rangle   
\nonumber \\
= && \exp \left (-i[(E_n -E_m)t + (l_n^2-l_m^2)Y(t)] \right ) \nonumber \\
  && \times \exp[-(l_n-l_m)^2 X(t)]\rho_{mn}(0) ,
\end{eqnarray}
where $E_n$ and $l_n$ are defined through
$H_{\rm IRHM}|n\rangle = E_n|n\rangle$ and $\sum_i S^z_i
|n\rangle =l_n|n\rangle$.
Furthermore, 
$X(t) \equiv \int_0^{t} F_R(s)ds $ and $Y(t) \equiv \int_0^{t} F_I(s)ds $ 
with $F_R(t) + i F_I(t) \equiv F(t)$.
This implies  that,
when states $|m\rangle$ and $|n\rangle$ have the same total spin $S_T$ and the
same z-component of the total spin $S^z_{T}$ ( i.e., $E_n=E_m$ and $l_n=l_m$),
 the  matrix elements $\rho_{mn} (t)= \rho_{mn}(0)$ and the states 
$|m\rangle$ and $|n\rangle$ belong to DFS.
Furthermore, the density matrix for two qubits at sites $i,j$ (obtained
from the groundstates of our SU(2) symmetric IHRM
Hamiltonian)
exhibits  time independence, that is,
$\rho_{ij}(t) = \rho_{ij}(0)$ with its elements given by Eq. (\ref{d_mat}).
This shows that 
  the dynamics of the system has no effect on the
 entanglement of the ground state of the system
which implies that these states are highly/maximally robust.
Consequently, our high $E^2_v$  RVB states constructed from 
the ground states of our IRHM 
 are also free of decoherence.
In future, using the above framework, we will consider the interesting case of
 dynamical evolution and decoherence
of a superposition of states with different 
 $l_n$ values.

\section{Conclusions.}
Although both spins in a two-spin singlet state 
{\nolinebreak$|\uparrow\downarrow\rangle-|\downarrow\uparrow\rangle$} are
monogamous (i.e., they cannot be entangled with any other spin),
by using a homogenized superposition of valence bond states (each of which
is a product of $N/2$
two-spin singlets), we managed to distribute entanglement
 efficiently 
such that any pair is maximally entangled with the rest of the RVB system
while concomitantly the constituent spins of the pair are completely unentangled
with each other.  Thus we get the converse of monogamy for a pair of spins!
Now, while total spin zero states are quite commonly ground states
(as shown by Lieb-Mattis theorem \cite{lieb}), it has not been recognized that one can 
generate high bipartite $E^2_v$ entanglement from such states.
 Our maximal $E^2_v$ RVB states are physically
realizable 
in systems such as 
infinite-range large $U/t$ Hubbard model and infinite-range hard-core
boson model with frustrated hopping \cite{ashvin} when they are at 
half-filling. 

Our high $E^2_v$ entanglement RVB states are free of decoherence when the
system interacts locally with optical phonons. Furthermore, the 
set of orthogonal ground states of our IRHM form a DFS and evolve unitarily
when all qubits are exposed to the same decoherening collective noise. 
 Thus our maximal $E^2_v$
RVB states are free of decoherence for two extreme types of coupling with the environment!
Our decoherence free RVB states can be used for
 error free quantum computation and quantum communication.

In summary, our maximal $E^2_v$ RVB states have led us to IRHM for their realization and IRHM
in turn generated maximal $E^2_v$ states as highly robust ground states.

\appendix

\section{Existence of homogeneous states for $N (\ge 8)$ qubit systems}

Let $|\psi_n\rangle$ be a set of linearly independent $S_T=0$ states with number of linearly 
independent states $n_L=\frac{N!}{(N/2)!(N/2 +1)!}$. Then the homogeneous ground state can be written as:
\begin{eqnarray}
|\Psi_{HGS}\rangle = \sum_{n=1}^{n_L} (\alpha_n  + i \beta_n) |\psi_n\rangle
\end{eqnarray} 
where $\alpha_n$ and $\beta_n$ are real. 
We have ${^N}C_2$ number of $\langle S^z_i S^z_j\rangle$ correlation functions appearing
in the expression for $E^2_v$.
For $N\geq 8$, we note that 
 ${^N}C_2 \leq  2n_L$; in fact, the ratio $2 n_L/{^N}C_2$ is unity for $N=8$ while for
$N>8$ the ratio monotonically
and rapidly
increases.
In order to maximize the system $E_v^2$, we must have $\langle S^z_i S^z_j\rangle$ independent
 of $i$ and $j$ and thus make the system homogeneous. Homogeneity condition
implies that the equality of the
${^N}C_2$ correlation functions $\langle S^z_i S^z_j\rangle$ 
produces
 ${^N}C_2 - 1$ independent equations.
Now, in $|\Psi_{HGS} \rangle$, there are only $2n_L-1$ independent $\alpha_n$ and $\beta_n$
 because of the normalization constraint.
Furthermore for $N \ge 8$, in the ${^N}C_2 -1$ equations involving $\langle S^z_i S^z_j\rangle$,
the number of independent coefficients is greater than or equal the number of equations, i.e.,
$2n_L-1 \geq {^N}C_2 -1$.
Thus there are enough number of unknowns to produce homogeneity. 

\section{Third-order perturbation for IRHM system coupled to local phonons}
\begin{figure}[]
\begin{center}
\includegraphics[width=3.5in,height=3.0in]{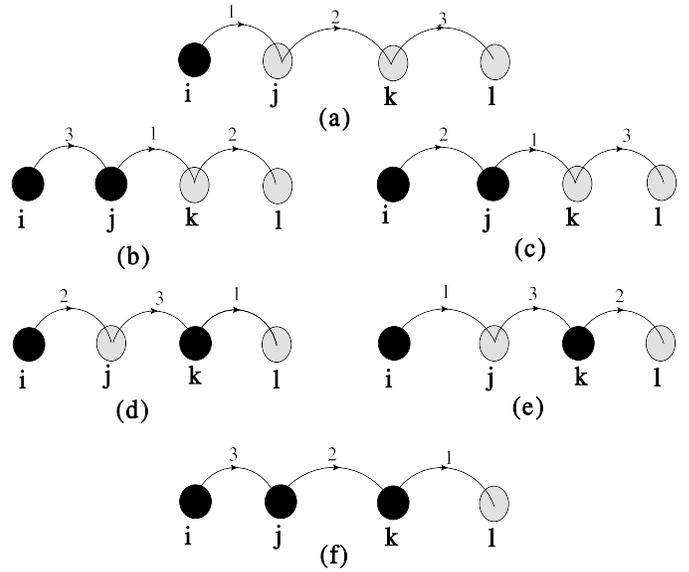}
\caption{ Open loop hopping processes contributing to effective hopping term $T_n^{li}$ in
third-order 
perturbation theory.
Here empty circles
correspond to sites with no particles while filled circles correspond 
to sites with hard-core-bosons. The numbers 1, 2, and 3 indicate the order of
hopping.
}
\label{fey1}
\end{center}
\end{figure}
\begin{figure}[]
\begin{center}
\includegraphics[width=1.5in,height=3.0in]{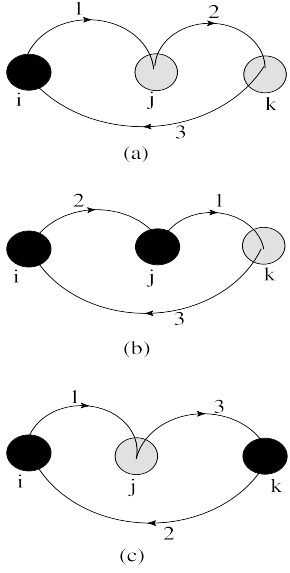}
\caption{Closed-loop hopping processes contributing to effective interaction term $V_n^i$ in
third-order 
perturbation theory.
Here filled (empty) circles 
correspond to sites with (without) hard-core-bosons. The numbers 1, 2, and 3 represent hopping sequence.}
\label{fey2}
\end{center}
\end{figure}
\begin{figure}[]
\begin{center}
\includegraphics[width=1.3in,height=3.0in]{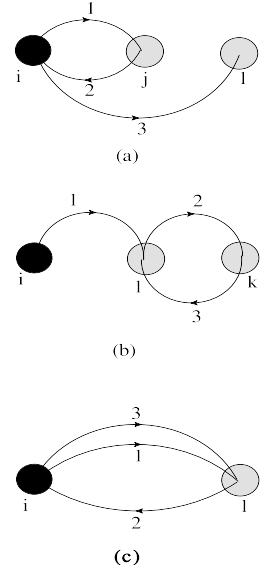}
\caption{Hopping processes (involving closed loops) contributing to effective hopping term $T_{Cn}^{li}$ in
third-order 
perturbation theory. Filled (empty) circles represent occupied (unoccupied) sites.
}
\label{fey3}
\end{center}
\end{figure}

 The total Hamiltonian $H_T$, involving interaction with local optical
phonons, is given by
\begin{eqnarray}
\!\!\!\! H_T = H_{\rm IRHM}+ g \omega_0 \sum_i S^z_i (a^{\dagger}_i + a_i) + \omega_0 \sum_i a^{\dagger}_i a_i ,
\label{Ham3}
\end{eqnarray}
where $a$ is the phonon destruction operator, $\omega_0$ is the optical phonon frequency, and $g$ is the
coupling strength. Now, we make the connection that the spin operators can be expressed in terms of
hard-core-boson (HCB) creation and destruction operators $b^{\dagger}$ and $b$, i.e., 
$b^{\dagger} = S^{+}$, $b = S^{-}$, and $b^{\dagger} b = S^{z} + 0.5$. We then
observe that conservation of $S^z_{Total}$ implies conservation of total number of HCB. 
The total Hamiltonian is then given by
\begin{eqnarray}
H &=&  J \sum_{i,j>i}[0.5 b^{\dagger}_i  b_{j} +{\rm H.c.} + n_in_j  ] 
\nonumber \\
          &&+ \omega_0 \sum_j a^{\dagger}_{j} a_j
          + g \omega_0 \sum_j n_j 
 (a_j +a^{\dagger}_j) , 
\end{eqnarray}
where
 $n_j \equiv b^{\dagger}_{j} b_j $.
Then we perform 
the well-known Lang-Firsov (LF) transformation \cite {lang,sdadys} on this
Hamiltonian. 
Under the LF transformation  given by $e^S H e^{-S} =H_0+H_1 $ with
$S= - g \sum_i n_i (a_i - a^{\dagger}_i)$, the operators
$b_j$ and $a_j$ transform like fermions  and phonons in the Holstein model.
This is due to the interesting 
 commutation properties of HCB given below: 
\begin{eqnarray}
[b_i,b_j]&=&[b_i,b^{\dagger}_j]= 0 , \textrm{ for } i \neq j , \nonumber\\
\{b_i,b^{\dagger}_i\}& = & 1 .
\label{commute}
\end{eqnarray}   
Next, the unperturbed Hamiltonian $H_0$ is identified to be 
\cite{sdadys}
\begin{eqnarray}
H_0 &=&  J \sum_{i,j>i}[(0.5 e^{-g^2}b^{\dagger}_i  b_{j} +{\rm H.c.}) + n_i n_j  ]
\nonumber \\
 && +\omega_0 \sum_j a^{\dagger}_j a_j 
      - g^2 \omega_0 \sum_j
n_j ,
\end{eqnarray}
while the perturbation $H_1$ is given by
\begin{eqnarray}
H_1
= J \sum_{i,j>i}[0.5 e^{-g^2}b^{\dagger}_i  b_{j} +{\rm H.c.}]
            \{\mathcal S^{{ij}^\dagger}_+ \mathcal S^{ij}_{-}-1\} ,
\end{eqnarray}
where $\mathcal S^{ij}_{\pm} = \textrm{exp}[\pm g(a_i - a_{j})]$ 
and $g^2 \omega_0 $
is the polaronic binding energy.\\
Since the unperturbed Hamiltonian $H_{0}$ does not contain any interaction
terms, we represent its eigenstates as
$|n,m\rangle\equiv|n\rangle_{hcb}\otimes|m\rangle_{ph}$ with the corresponding
 eigenenergies $E_{n,m}=E_{n}^{hcb}+E_{m}^{ph}$.
On observing that $\langle0,0|H_{1}|0,0\rangle=0$,
we obtain the next relevant second-order perturbation term \cite{sdadys}
\begin{eqnarray}
E^{(2)}=\sum_{n,m}{{\langle0,0|H_{1}|n,m\rangle\langle n,m|H_{1}|0,0\rangle}\over{E_{0,0}-E_{n,m}}} .
\end{eqnarray}
For strong coupling ($g>1$) and non-adiabatic ($J/\omega_0 \leq 1$) conditions,
on noting that 
$E_{m}^{ph}-E_{0}^{ph}$ is a positive integral multiple of $\omega_{0}$  and 
$E_{n}^{hcb}-E_{0}^{hcb}
\sim Je^{-g^2} \ll \omega_0$, we 
get the following second-order term \cite{sdys}
\begin{eqnarray}
H^{(2)} \! &=& \! \sum_{i, j > i } 
 \left [(0.5 J_{\perp}^{(2)}
b^{\dagger}_i
 b_j +{\rm H.c.}) + 
 J_{\parallel}^{(2)}
 n_in_j \right ] ,
\nonumber \\
\label{H_eff}
\end{eqnarray}
where 
\begin{eqnarray}
\!\!\!\!\!\!\!\!\!\!\!\! J_{\perp}^{(2)} \equiv  - (N-2) f_1 (g) \frac{J^2 e^{-2g^2}}{2 \omega_0}
 \sim -(N-2) \frac{J^2e^{-g^2}}{2g^2 \omega_0} ,
 \end{eqnarray}
\begin{eqnarray}
 J_{\parallel}^{(2)} \equiv 
[2f_1 (g)+f_2(g)]\frac{J^2 e^{-2g^2}}{{2 \omega_0}} 
\sim \frac{J^2 }{{4g^2 \omega_0}}, 
 \end{eqnarray}
 with
$f_1(g) \equiv \sum^{\infty}_{n=1} g^{2n}/(n!n)$
 and
$f_2(g) \equiv \sum^{\infty}_{n=1}\sum^{\infty}_{m=1} g^{2(n+m)}/[n!m!(n+m)]$. 
The small parameter for our perturbation theory is $t/(g \omega_0)$.
We now make the important observation
that the eigenstates of the effective Hamiltonian $H_{0}+H^{(2)}$ 
 are identical to
those of the original Hamiltonian $H_{\rm IRHM}$ in Eq. (\ref{H_gen}) because $\sum_{i, j > i } 
 (S_i^{z} S_j^{z})$ and $ H_{\rm IRHM}$ commute.

Next, we will show that the third-order perturbation theory also produces a term that has the same
eigenstates as IRHM.
To this end, we obtain the following third-order perturbation term in the effective Hamiltonian:\\
\begin{eqnarray}
\!\!\!H^{(3)}\!=\sum_{m\neq 0,n \neq 0}\!\!\!\! \frac{{_{ph}}\!\langle0|H_{1}|m\rangle_{ph}
~{_{ph}}\!\langle m|H_{1}|n\rangle_{ph}
~{_{ph}}\!\langle n|H_{1}|0\rangle_{ph}}
{{\Delta E_m}{\Delta E_n}}  . 
\nonumber \\ 
\label{H3}
\end{eqnarray}
Here $\Delta E_{m}= E_{m}^{ph}-E_{0}^{ph}$.
Evaluation of $H^{(3)}$ leads to various hopping terms and interaction terms.
\begin{eqnarray}
H^{(3)}=\sum_{i,l\neq i} \left [\sum_{n=1}^{6} t_n T_n^{li} 
   + \sum_{n=1}^3 t_{cn} T_{Cn}^{li} \right ] + \sum_i \sum_{n=1}^3 v_n V^i_n , 
\nonumber \\ 
\label{tvtc}
\end{eqnarray}
where $t_n \sim (J^3 e^{-g^2})/(g^2 \omega_0)^2$,
 $t_{cn} \sim J^3 e^{-g^2}/(g\omega_0)^2$, and  $v_n \sim J^3 /(g^2 \omega_0)^2$
(as will be explained later).
We will demonstrate below that  $H^{(3)}$ is of the following form
\begin{eqnarray}
 H^{(3)} =\sum_{i,l > i} \left [ T(\sum_k n_k) b^{\dagger}_l b_i + {\rm H.c.} \right ] +  \sum_{i} V(\sum_k n_k) n_i  . 
\nonumber \\ 
\label{H3_form}
\end{eqnarray}
where $T$ and $V$ are functions of the total number operator $\sum_k n_k$.
 Since the IRHM commutes with the total number operator, $H^{(3)}$ has the same eigenstates as IRHM! 

 There are six open-loop hopping processes $T_n^{li}$ depicted in Fig. \ref{fey1}.  We analyze
them sequentially below.
\begin{eqnarray}
\!\!\!\!\!\!\!\!\!\!\!\!\!\!\!\!T_1^{li}
 &=& \sum_{k \neq i,l,j}\sum_{j \neq i,l} b^{\dagger}_l b_k b^{\dagger}_k b_j b^{\dagger}_j b_i 
\nonumber \\
 &=& \sum_{k \neq i,l,j} (1-b^{\dagger}_k b_k) \sum_{j \neq i, l} (1-b^{\dagger}_j b_j) b^{\dagger}_l b_i 
\nonumber \\
&=& \left [\sum_{k \neq i,l} (1-b^{\dagger}_k b_k) -1 \right ] \left [ \sum_{j \neq i, l} 
(1-b^{\dagger}_j b_j)\right ] b^{\dagger}_l b_i 
\nonumber \\
&=&\left [ \sum_{k \neq i,l} (1-b^{\dagger}_k b_k)-1 \right ] \left [ (N-2) - 
\sum_{j \neq l} b^{\dagger}_j b_j \right ] b^{\dagger}_l b_i
\nonumber \\
&=& \left [\sum_{k \neq i,l} (1-b^{\dagger}_k b_k)-1 \right ] \left [(N-1) - \sum_{j } b^{\dagger}_j b_j \right ] b^{\dagger}_l b_i
\nonumber \\
&=& \left [(N-1) - \sum_{j } b^{\dagger}_j b_j \right ] 
\left [\sum_{k \neq i,l} (1-b^{\dagger}_k b_k)-1 \right ]b^{\dagger}_l b_i
\nonumber \\
&=&\left [(N-1) - \sum_{j } b^{\dagger}_j b_j \right ] 
\left [(N-2) -\sum_{k } b^{\dagger}_k b_k \right ]b^{\dagger}_l b_i .
\nonumber \\
\end{eqnarray}
The second hopping
process $T_2^{li}$ in Fig. \ref{fey1} (b)  is given by 
\begin{eqnarray}
\!\!\!\!\!\!\!\!\!\!\!\!\!\!\!\!T_2^{li}
 &=& \sum_{k \neq i,l,j}\sum_{j \neq i,l} b^{\dagger}_j b_i b^{\dagger}_l b_k b^{\dagger}_k b_j 
\nonumber \\
 &=& \sum_{k \neq i,l,j} (1-b^{\dagger}_k b_k) \sum_{j \neq i, l} b^{\dagger}_j b_j b^{\dagger}_l b_i 
\nonumber \\
&=& \sum_{k \neq i,l} (1-b^{\dagger}_k b_k) \sum_{j \neq i, l} 
b^{\dagger}_j b_j b^{\dagger}_l b_i 
\nonumber \\
&=&\sum_{k \neq i,l} (1-b^{\dagger}_k b_k) \left [ 
\sum_{j } b^{\dagger}_j b_j -1 \right ] b^{\dagger}_l b_i
\nonumber \\
&=& \left [ 
\sum_{j } b^{\dagger}_j b_j -1 \right ]\left [ (N-1)-\sum_{k}b^{\dagger}_k b_k) \right ] b^{\dagger}_l b_i .
\end{eqnarray}
The hopping process $T_3^{li}$ in Fig. \ref{fey1} (c)  
is expressed as $T_3 ^{li}=\sum_{k \neq i,l,j}\sum_{j \neq i,l} b^{\dagger}_l b_k b^{\dagger}_j b_ib^{\dagger}_k b_j
 =T_2^{li}$.
The fourth hopping process $T_4^{li}$  in Fig. \ref{fey1} (d) is obtained as follows.
\begin{eqnarray}
\!\!\!\!\!\!\!\!\!\!\!\!\!\!\!\!T_4^{li}
 &=& \sum_{j \neq i,l, k}\sum_{k \neq i,l} b^{\dagger}_k b_j b^{\dagger}_j b_i b^{\dagger}_l b_k 
\nonumber \\
 &=& \sum_{j \neq i,l,k} (1-b^{\dagger}_j b_j) \sum_{k \neq i, l} b^{\dagger}_k b_k b^{\dagger}_l b_i 
\nonumber \\
&=& T_2^{li}  .
\end{eqnarray}
The hopping process $T_5^{li}$  in Fig. \ref{fey1} (e) yields
 $T_5^{li} =\sum_{j \neq i,l, k}\sum_{k \neq i,l} b^{\dagger}_k b_j  b^{\dagger}_l b_k b^{\dagger}_j b_i 
=T_4^{li}$.
 We  analyze below the last hopping 
process $T_6^{li}$ in Fig. \ref{fey1} (f).
\begin{eqnarray}
\!\!\!\!\!\!\!\!\!\!\!\!\!\!\!\!T_6^{li}
 &=& \sum_{k \neq i,l,j}\sum_{j \neq i,l} b^{\dagger}_j b_i b^{\dagger}_k b_j b^{\dagger}_l b_k 
\nonumber \\
 &=& \sum_{k \neq i,l,j} b^{\dagger}_k b_k \sum_{j \neq i, l} b^{\dagger}_j b_j b^{\dagger}_l b_i 
\nonumber \\
&=& \left [\sum_{k \neq i,l} b^{\dagger}_k b_k - 1 \right ] \sum_{j \neq i, l} 
b^{\dagger}_j b_j b^{\dagger}_l b_i 
\nonumber \\
&=&\left [\sum_{k \neq i,l} b^{\dagger}_k b_k - 1 \right ]\left [ 
\sum_{j } b^{\dagger}_j b_j -1 \right ] b^{\dagger}_l b_i
\nonumber \\
&=& \left [ 
\sum_{j } b^{\dagger}_j b_j -1 \right ]\left [ \sum_{k}b^{\dagger}_k b_k-2 \right ] b^{\dagger}_l b_i  .
\end{eqnarray}

We will now deal with closed-loop hopping processes such as those in Fig \ref{fey2}.
These lead to effective interactions. The process $V_1^i$ in Fig. \ref{fey2} (a),  obtained from
Fig. \ref{fey1} (a) by setting $l=i$, is given as follows. 
\begin{eqnarray}
\!\!\!\!\!\!\!\!\!\!\!\!\!\!\!\!V_1^i
 &=& \sum_{k \neq i,j}\sum_{j \neq i} b^{\dagger}_i b_k b^{\dagger}_k b_j b^{\dagger}_j b_i 
\nonumber \\
 &=& \sum_{k \neq i,j} (1-b^{\dagger}_k b_k) \sum_{j \neq i} (1-b^{\dagger}_j b_j) b^{\dagger}_i b_i 
\nonumber \\
&=& \left [\sum_{k \neq i} (1-b^{\dagger}_k b_k) -1 \right ] \left [ \sum_{j \neq i} 
(1-b^{\dagger}_j b_j)\right ] b^{\dagger}_i b_i 
\nonumber \\
&=&\left [(N) - \sum_{j } b^{\dagger}_j b_j \right ] 
\left [(N-1) -\sum_{k } b^{\dagger}_k b_k \right ]b^{\dagger}_i b_i .
\end{eqnarray}
Next, the hopping process $V_2^i$ corresponding to closed loop in Fig. \ref{fey2} (b)
 is obtained from Fig. \ref{fey1} (c) by taking $l=i$.
\begin{eqnarray}
\!\!\!\!\!\!\!\!\!\!\!\!\!\!\!\!V_2^i
 &=& \sum_{k \neq i,j}\sum_{j \neq i} b^{\dagger}_i b_k b^{\dagger}_j b_i b^{\dagger}_k b_j 
\nonumber \\
 &=& \sum_{k \neq i,j} (1-b^{\dagger}_k b_k) \sum_{j \neq i} b^{\dagger}_j b_j b^{\dagger}_i b_i 
\nonumber \\
&=& \sum_{k \neq i} (1-b^{\dagger}_k b_k) \sum_{j \neq i} 
b^{\dagger}_j b_j b^{\dagger}_i b_i 
\nonumber \\
&=&\sum_{k \neq i} (1-b^{\dagger}_k b_k) \left [ 
\sum_{j } b^{\dagger}_j b_j -1 \right ] b^{\dagger}_i b_i
\nonumber \\
&=& \left [ 
\sum_{j } b^{\dagger}_j b_j -1 \right ]\left [ (N)-\sum_{k}b^{\dagger}_k b_k) \right ] b^{\dagger}_i b_i .
\end{eqnarray}
Lastly, the hopping $V_3^i$ [depicted by the closed loop in Fig. \ref{fey2} (c)]
 is obtained from Fig. \ref{fey1} (e) by setting $l=i$.
\begin{eqnarray}
\!\!\!\!\!\!\!\!\!\!\!\!\!\!\!\!V_3^i
 &=& \sum_{j \neq i, k}\sum_{k \neq i} b^{\dagger}_k b_j b^{\dagger}_i b_k b^{\dagger}_j b_i
\nonumber \\
 &=& \sum_{j \neq i,k} (1-b^{\dagger}_j b_j) \sum_{k \neq i} b^{\dagger}_k b_k b^{\dagger}_i b_i 
\nonumber \\
&=& V_2^i  .
\end{eqnarray}

Finally, we consider Figs. \ref{fey3} (a), (b), and (c) 
 which deal with effective hopping terms $T_{Cn}^{li}$ involving closed loops. The effective hopping
 term $T_{C1}^{li}$,
corresponding to Fig. \ref{fey3} (a), is obtained by setting $k=i$ in  Fig. \ref{fey1} (a):
\begin{eqnarray}
\!\!\!\!\!\!\!\!\!\!\!\!\!\!\!\!T_{C1}^{li}
 &=& \sum_{j \neq i, l} b^{\dagger}_l b_i b^{\dagger}_i b_j b^{\dagger}_j b_i 
\nonumber \\
 &=&  \sum_{j \neq i, l} (1-b^{\dagger}_j b_j) b^{\dagger}_l b_i 
\nonumber \\
&=& \left [ (N-2) - 
\sum_{j \neq l} b^{\dagger}_j b_j \right ] b^{\dagger}_l b_i
\nonumber \\
&=&  \left [(N-1) - \sum_{j } b^{\dagger}_j b_j \right ] b^{\dagger}_l b_i .
\end{eqnarray}
To obtain the effective hopping term $T_{C2}^{li}$
corresponding to Fig. \ref{fey3} (b), we take $j=l$ in  Fig. \ref{fey1} (a):
\begin{eqnarray}
\!\!\!\!\!\!\!\!\!\!\!\!\!\!\!\!T_{C2}^{li}
 &=& \sum_{k \neq i,l}
b^{\dagger}_l b_k b^{\dagger}_k b_l b^{\dagger}_l b_i 
\nonumber \\
 &=& \sum_{k \neq i,l} (1-b^{\dagger}_k b_k)  b^{\dagger}_l b_i 
\nonumber \\
&=& \left [ (N-2) - 
\sum_{k \neq l} b^{\dagger}_k b_k \right ] b^{\dagger}_l b_i
\nonumber \\
&=& \left [(N-1) - \sum_{k } b^{\dagger}_k b_k \right ] b^{\dagger}_l b_i 
\nonumber \\
&=& T_{C1}^{li} .
\end{eqnarray}
The effective hopping term $T_{C3}^{li}$ depicted in Fig. \ref{fey3} (c) 
[upon setting $k=i$ and $j=l$ in  Fig. \ref{fey1} (a)] is given by
\begin{eqnarray}
\!\!\!\!\!\!\!\!\!\!\!\!\!\!\!\!T_{C3}^{li}
 = b^{\dagger}_l b_i b^{\dagger}_i b_l b^{\dagger}_l b_i 
= b^{\dagger}_l b_i .
\end{eqnarray}

Thus we have shown that $H^{(3)} $ contains effective hopping terms 
($\sum_{i,l>i} \left [ T(\sum_k n_k) b^{\dagger}_l b_i + {\rm H.c.} \right ]$)
 and effective interaction terms ($\sum_{i} V(\sum_k n_k) n_i$). Since 
 $T$ and $V$ are functions of the total number operator,
 $H^{(3)}$ and IRHM have the same eigenstates. These arguments can be extended to even
higher-order perturbation theory to show that the effective Hamiltonian (after taking all
orders of perturbation into account) will give the same eigenstates as IRHM!

\begin{figure}[]
\begin{center}
\includegraphics[width=1.5in,height=3.0in]{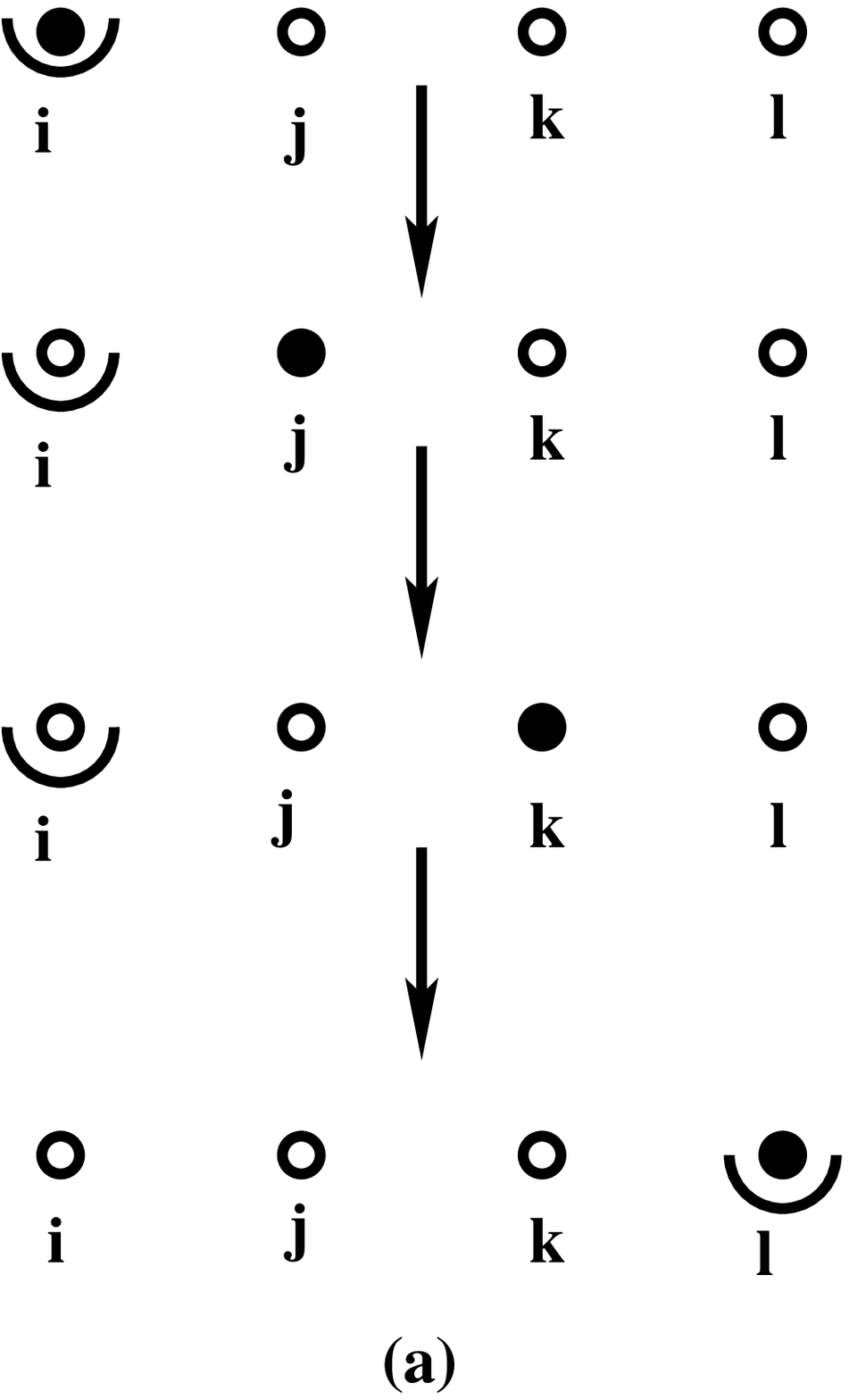}\\
\vspace{.3cm}
\includegraphics[width=2.5in,height=2.0in]{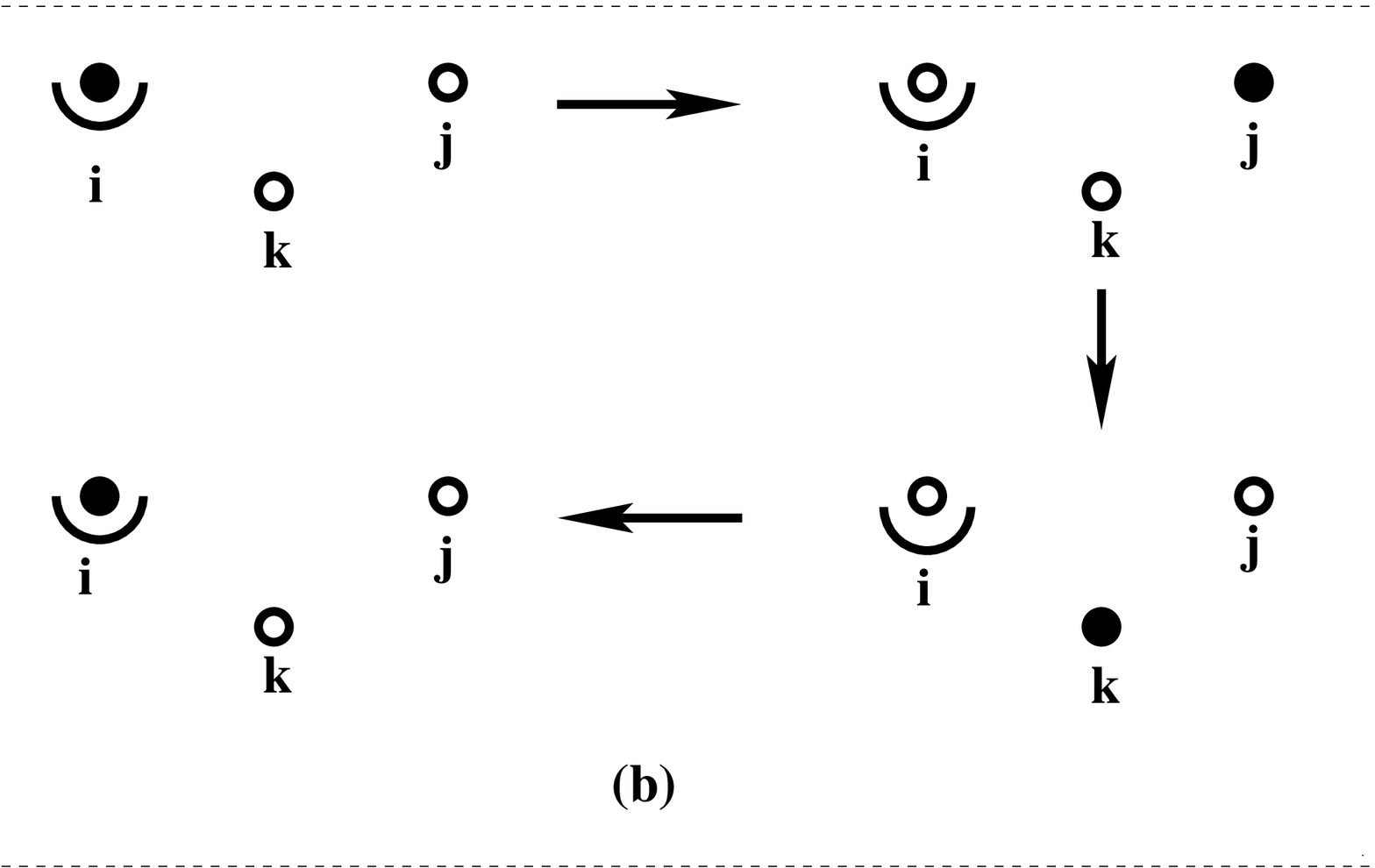}\\
\vspace{.3cm}
\includegraphics[width=2.5in,height=2.0in]{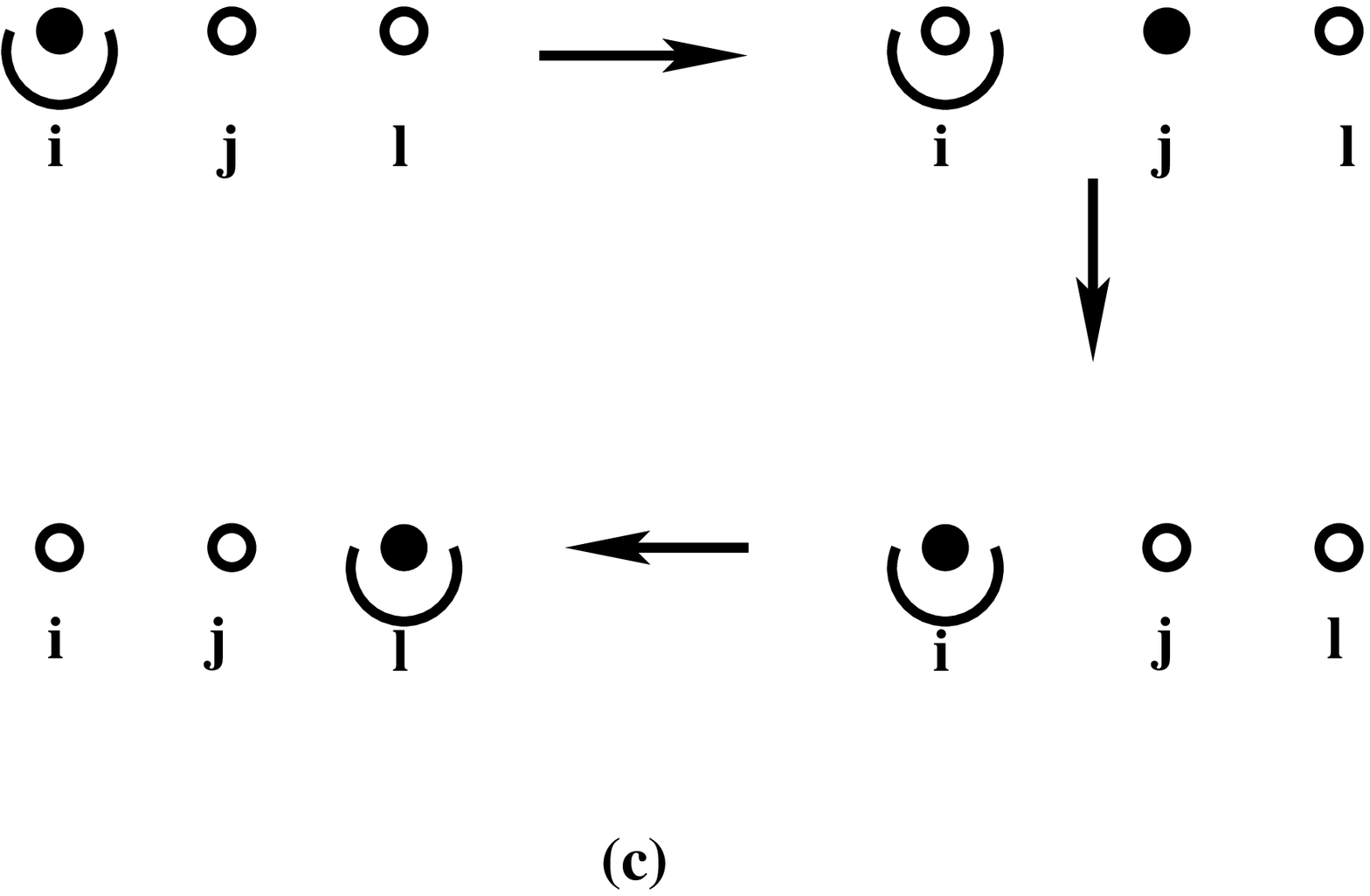}
\caption{Schematic diagrams (a), (b), and (c), corresponding
to the hopping processes depicted in Fig. \ref{fey1} (a), Fig. \ref{fey2} (a), 
and Fig. \ref{fey3} (a) respectively,  yield coefficients
$t_n$, $v_n$, and $t_{cn}$ respectively.  The intermediate states give the typical dominant
contributions. Here empty circles correspond to empty sites, while filled circles
 indicate particle positions. Parabolic curve
at a site depicts full distortion at that site with
corresponding energy $-g^2 \omega_0$ ($+g^2 \omega_0$) if the hard-core-boson is present
(absent) at that site.
}
\label{schemat}
\end{center}
\end{figure}
We will now explain the expressions for the coefficients $t_n$, $v_n$, and $t_{cn}$
in Eq. (\ref{tvtc}), 
 obtained from third-order
perturbation theory, using typical schematic
diagrams shown in Fig. \ref{schemat} \cite{srsypbl}. We consider two distinct time scales
associated with  hopping processes between two sites:
(i) $\sim 1/(Je^{-g^2})$ corresponding to either full distortion at a site
to form a small polaronic potential well (of energy $-g^2 \omega_0$) or
full relaxation from the small polaronic distortion and (ii) $\sim 1/J$
related to negligible distortion/relaxation at a site. 
The coefficient $t_n$ corresponds
to the typical dominant distortion processes shown schematically in Fig. \ref{schemat} (a)
with the pertinent typical  hopping processes being depicted in
Fig. \ref{fey1} (a). In Fig. \ref{schemat} (a), after the HCB hops
 away from the initial site, the intermediate states
have the same distortion as the initial state. Next,
when the HCB hops to its final site there is a distortion
at this final site with a concomitant relaxation at the initial site.
Hence the contribution to the coefficient $t_n$ becomes
$J/(2 g^2 \omega_0) \times J/(2 g^2 \omega_0) \times J e^{-g^2} \sim J^3 e^{-g^2}/(g^2 \omega_0)^2$.
As regards coefficient $v_n$, it can be deduced based on the typical dominant 
hopping-cum-distortion 
processes depicted in Fig. \ref{schemat} (b) which typifies the hopping processes in Fig. \ref{fey2} (a).
In Fig. \ref{schemat} (b), when the particle hops to different sites and reaches finally the initial site,
there is no change in distortion at any site. Hence $V_n$ can be estimated to be
$J/(2 g^2 \omega_0) \times J/(2 g^2 \omega_0) \times J  \sim J^3 /(g^2 \omega_0)^2$.
Lastly, we obtain the coefficient $t_{cn}$ by considering the 
typical dominant diagram in Fig. \ref{schemat} (c)
corresponding to the typical process in Fig. \ref{fey3} (a). In Fig. \ref{schemat} (c),
where the first intermediate state depicts the particle hopping but leaving the distortion unchanged,
we get a contribution $J/(2g^2 \omega_0)$; for the next intermediate state, where the HCB returns to
the initial site, the initial site has to undergo a slight relaxation (involving absorbing a phonon
so as to yield a non-zero denominator in the perturbation theory) leading to the contribution $J/\omega_0$;
and lastly, when the HCB hops to the final site, there is a distortion
at the final site with a simultaneous relaxation at the initial site thereby producing
a contribution $J e^{-g^2}$. Thus we calculate $v_n$ to be
 $J/(2g^2 \omega_0) \times J/\omega_0 \times J e^{-g^2} \sim J^3 e^{-g^2}/(g \omega_0)^2$ \cite{sy1}.


\begin{thebibliography}{9}
\bibitem{note} In Ref. \onlinecite{sen} it was pointed out that the ``RVB liquid'' on a 
$4 \times 4$ square lattice gives $p > 1/3$ for periodic boundary conditions while
it yields $p < 1/3$ in the interior of a sample with open boundary conditions.
\end{thebibliography}
\end{document}